\documentclass{phostproc}

\title{Influence of structural discontinuities present in the core of red-giant stars on the observed mixed-mode pattern and characterization of their properties}
\author{Mathieu Vrard,$^{1}$ 
        Margarida S. Cunha$^{1}$}

\affiliation{$^{1}$ Instituto de Astrof\'isica e Ci\^encias do Espa\c{c}o, Universidade do Porto, CAUP, Rua das Estrelas, 4150-762 Porto, Portugal}

\shorttitle{Discontinuities in the core of red-giant stars}
\shortauthors{Mathieu Vrard \& Margarida S. Cunha}

\newcommand{\ind}[1]{_{\mathrm{#1}}}
\newcommand{\diff}{\mathrm{d}}
\newcommand{\e}{e}
\newcommand{\arccot}{arcCot}

\def\Kepler{\emph{Kepler}}
\def\numax{\nu\ind{max}}
\def\nmax{n\ind{max}}
\def\Dnu{\Delta\nu}
\def\deltapi{\Delta\Pi_{1}}
\def\eps_g{\varepsilon\ind{g}}
\def\epsilonp{\varepsilon\ind{p}}
\def\np{n\ind{p}}
\def\Brunt{N}
\def\Lamb{S\ind{\ell}}
\def\Radreg{\mathcal{R}}
\def\RadP{\mathcal{P}}
\def\Phasep{\theta_p}
\def\Phaseg{\theta_g}
\def\waveN{k_r}
\def\taufreq{\tau}
\def\red{x_{\nu}}

\def\coupling{\varphi}
\def\glitch{\Phi}

\def\Amp{\mathcal{A}}
\def\Period{\omega_g^{*}}
\def\cavity{\omega_g^{r}}
\def\rescavity{\omega_g}
\def\width{\Delta_g}

\abs{The space-borne missions CoRoT and $\Kepler$ have provided seismic data of unprecedented quality. Among the observed stars, red giants show a complex oscillation pattern exhibiting pressure modes as well as mixed modes. The latter carry information on the radiative region properties of these stars. The very high precision of $\Kepler$ data provide enough accuracy to decipher the complex structure of the mixed-mode pattern and, therefore, deduce precise information on the structure of the stellar core. In this work, we studied the precise influence of the core structural discontinuities on the mixed-mode pattern. These phenomena have indeed an influence on the gravity waves that propagate inside the core of the star and, therefore, on the mixed-mode pattern. In this study, we aim to investigate the impact of core structural discontinuities on the observed mixed-mode pattern and measure their properties for several red giants. We identified several objects showing evidence of deviations in their mixed-mode frequency pattern that are characteristic of core structural discontinuities. We fitted these deviations and showed that they likely correspond to the influence of the inner convective core of these stars.}

\begin{document}

\maketitle

\section{Introduction}

Since the launch of the space-borne photometric missions CoRoT \citep{2006ESASP1306...33B} and $\Kepler$ \citep{2010Sci...327..977B}, many important studies have been carried out. Among the observed stars, red giants have shown particularly complex spectra, exhibiting pressure modes as well as mixed modes \citep{2009Natur.459..398D}. The latter are the results of modes behaving as gravity modes in the core of the star and as pressure modes in their envelope thus allowing to obtain information on their stellar core. They were used to distinguish core-helium burning stars (clump stars) from hydrogen-shell burning red-giant branch stars \citep[RGB; ][]{2011Sci...332..205B,2011Natur.471..608B,2012A&A...540A.143M} and also provide observational constraints on the stellar core rotation \citep{2012Natur.481...55B,2012A&A...548A..10M}.

The pressure modes present in red-giant star spectra result from acoustic waves stochastically excited by convection in the outer layers of the star. The observed pressure mode pattern has been depicted in a canonical form, called the universal red-giant oscillation pattern \citep{2011A&A...525L...9M}. This canonical form describes the regularity of the pressure mode pattern characterized by two quantities: the frequency $\numax$ of maximum oscillation and the mean frequency difference $\Dnu$ between consecutive pressure modes of same angular degree. We can consider the observed pattern to be the translation of the second-order asymptotic pattern described in \citet{1980ApJS...43..469T} at low radial order \citep{2013A&A...550A.126M}. 

The mixed modes show a more complicated pattern than pressure modes and gravity modes, which, for the latter, are evenly spaced in period. However, their oscillation pattern can  be asymptotically described \citep{Unno1989,2012A&A...540A.143M,2014MNRAS.444.3622J} and the deviations from this pattern can, therefore, be characterized. 

It has long been predicted that structural discontinuities exist in stellar interiors and that they affect the observed solar-like oscillations by inducing regular deviations from the classical mode pattern \citep[e.g. ][]{1990LNP...367..283G}. The influence of structural discontinuities, usually called glitches, on the pressure mode pattern was also investigated for red-giant stars \citep{2014MNRAS.tmp..576B,2014MNRAS.445.3685C} and observed in $\Kepler$ data \citep{2010A&A...520L...6M,2015A&A...579A..83C,2015A&A...579A..84V}. However, few studies investigated the influence of structural discontinuities on the gravity waves for the same type of objects. These type of studies were mainly done for gravity-mode pulsators, like white dwarfs \citep[e.g.][]{1991ApJ...378..326W,1992ApJS...80..369B}, $\gamma$-Doradus stars \citep[e.g. ][]{2008MNRAS.386.1487M} and hot B subdwarfs stars \citep{2000ApJS..131..223C,2018MNRAS.474.4709K} which are essentially the cores of previous red giants. Only one recent study transposed the previous work done on the aforementioned gravity-mode pulsators to the inner g-mode resonant cavity of red-giant stars and, thus, deduced the influence of the structural discontinuities located in their deeper layer on their mixed-mode pattern \citep{2015ApJ...805..127C}. The signature of these discontinuities was then claimed to have been discovered by \citet{2015A&A...584A..50M} but the precise characterization of these signatures was not performed. The aim of this work is to use the analytic model describing the influence of glitches on the mixed-mode pattern that was derived by \citet[][in preparation]{2015ApJ...805..127C} and use it to characterize the discontinuity characteristics. First, we will describe this analytical model and adapt it to estimate the frequency position of the mixed modes as a function of the discontinuity characteristics. Second, we will use this analytical description to measure the glitch characteristics in several red-giant clump stars observed with $\Kepler$.

\section{Description of the glitch influence on the red-giant mixed-mode pattern}

\subsection{Analytical description}

Asymptotically, gravity modes are approximately evenly spaced in period following the asymptotic period spacing $\Delta\Pi_{\ell}$. This asymptotic value is defined by the integration of the Brunt-Väisäla radial profile $\Brunt$ inside the radiative inner regions $\Radreg$. For $\ell = 1$ modes, it writes

\begin{equation}
   \deltapi = \frac{2\pi^{2}}{\sqrt{2}}\left(\int_{\mathcal{\Radreg}}
   \frac{\Brunt}{r}\;  \diff r\right)^{-1}.
    \label{Brunt_equation}
\end{equation}
Its value is related to the size of the inner radiative region \citep{2013EPJWC..4303002M}.
\newline

As stated before, gravity waves propagate only in regularly stratified medium, which correspond here to the stellar radiative core. Consequently, gravity modes will not be observed directly in red-giant star spectra. However, in red-giant stars, a coupling occur between gravity and pressure waves giving rise to the so-called mixed-modes. An asymptotic relation describing the mixed modes behavior was provided by \citet{Shibahashi1979} and \citet{Unno1989}. In this framework, eigenfrequencies are derived from an implicit equation relating the coupling of the p and g waves through an evanescent region \citep[Eq. (16.50) of][]{Unno1989}:

\begin{equation}\label{asympt_coupling}
  \tan \Phasep
  =
  q
  \tan \Phaseg,
\end{equation}
where $\Phasep$ and $\Phaseg$ are the p- and g-wave phases. The dimensionless coefficient q corresponds to the coupling between the p- and g-waves and measures the level of mixture of the p and g phases. Following \citet{Unno1989}, the p- and g-wave phases can be written as, for $\ell = 1$ modes:

\begin{equation}\label{PhaseP_obs}
     \Phasep = \pi \left(\frac{\nu-\nu_{n,\ell=1}}{\Dnu} \right),
\end{equation}
\begin{equation}\label{PhaseG_obs}
     \Phaseg = \pi\left(\frac{1}{\nu\deltapi} - \eps_g \right),
\end{equation}
where $\nu$ corresponds to the mixed-mode frequencies, $\nu_{n,\ell=1}$ represents the pure pressure $\ell = 1$ modes and $\eps_g$ is the gravity offset.

In the case of the JWKB approximation applied to non-radial adiabatic oscillations, $\Phasep$ and $\Phaseg$ can also be express as the following \citep{Shibahashi1979,Unno1989}: 

\begin{equation}\label{Description_PhaseP}
  \Phasep
  =
  \int_{\mathcal{\RadP}}   \waveN^{2}\;  \diff r
  \end{equation}

\begin{equation}\label{Description_PhaseG}
  \Phaseg
  =
  \int_{\mathcal{\Radreg}}   \waveN^{2}\;  \diff r,
\end{equation}
where $\RadP$ correspond to the pressure waves resonant cavity region and $\waveN$ is the radial wavenumber. This quantity can be approximated by the following expression:

\begin{equation}\label{Description_PhaseG}
  \waveN^{2}
  \approx \frac{\omega^{2}}{c_s^{2}} \left(\frac{\Lamb^{2}}{\omega^{2}} - 1 \right)
  \left(\frac{\Brunt^{2}}{\omega^{2}} - 1 \right),
\end{equation}
where $\omega$ is the wave angular frequency, $c_s$ is the sound speed and $\Lamb$ is the Lamb frequency for modes of degree $\ell$, expressed as $\Lamb = \sqrt{\ell(\ell+1)}c_s/r$.
\newline

To understand the impact of a glitch on the oscillation frequencies, we will consider a single discontinuity appearing at a specific position in radius in the buoyancy frequency (hereafter named $r^{*}$). At first, we model the glitch using a Gaussian-like function which we defined as the following for $\ell = 1$ modes: 

\begin{equation}\label{Description_PhaseG}
  \Brunt^{2} = \Brunt_0^{2} \left[1 + \frac{\Amp_G}{\sqrt{2\pi}\width} \left(\frac{r}{\sqrt{2}\Brunt_0}\right)^{\frac{1}{2}}  \exp \left(\frac{-(\cavity-\Period)^{2}}{2\width^{2}} \right)  \right],
\end{equation}
where $\Brunt_0$ is the glitch-free buoyancy frequency. The constants $\Amp_G$ and $\width$ measure respectively the amplitude and width of the glitch while the parameters $\cavity$ and $\Period$ correspond respectively to the buoyancy depth and the buoyancy depth at the glitch position. 

The buoyancy depth of a specific discontinuity corresponds to the position of the glitch in the considered resonant cavity as seen by the gravity waves. By defining $r_1$ and $r_2$ as the lower and upper turning points, respectively, that define the propagation cavity of the g-waves, the buoyancy depth can be expressed as: $\cavity = \sqrt{2} \int_{r}^{r_2} \frac{\Brunt}{r}\;  \diff r$. This quantity can be related to the gravity-mode period spacing ($\deltapi$) through the total buoyancy radius \citep{1980ApJS...43..469T}:

\begin{equation}\label{Relation_deltapi_buoyancyradius}
  \rescavity \approx \frac{2\pi^{2}}{\deltapi} = \cavity + \sqrt{2} \int_{r_1}^{r} \frac{\Brunt}{r}\;  \diff r.
\end{equation}

The buoyancy depth at the glitch position will be described as $\Period = \sqrt{2} \int_{r^{*}}^{r_2} \frac{\Brunt}{r}\;  \diff r$. The glitch is therefore defined by three parameters: $\Amp_G$, $\width$ and $\Period$.
\newline

The mode frequencies are determined by finding which oscillation eigenfrequencies are allowed by the boundary conditions that are imposed to the problem \citep[e.g.][]{2007AN....328..273G,2015ApJ...805..127C}. The work of \citet{2015ApJ...805..127C} demonstrated that the eigenvalue condition in the presence of mode coupling and a glitch is given by:

\begin{equation}\label{Equation_deltapi_glitch}
  \int_{r_1}^{r_2} \waveN\;  \diff r = \pi\left(n - \frac{1}{2} \right) - \coupling - \glitch,
\end{equation}
where $n$ is a positive integer, $\coupling \approx \arctan\left(\frac{q}{\tan(\Phasep)} \right)$ is the coupling phase and $\glitch$ represents the glitch frequency-dependent phase. The $\coupling$ and $\glitch$ parameters correspond to, respectively, the coupling and glitch influence on the mode frequencies.

If we assume that the coupling and glitch influence correspond to small perturbations, then we can write $\waveN = \waveN^{0} + \delta_k$ where $\waveN^{0}$ corresponds to the radial wavenumber without accounting for any glitch or coupling and $\delta_k$ is the perturbation due to the coupling and glitches. Since without glitch and coupling, it has been shown that the eigenvalue conditions translates to $\int_{r_1}^{r_2} \waveN^{0}\;  \diff r = \pi\left(n - \frac{1}{2} \right)$, we can therefore write:

\begin{equation}\label{Equation_glitch}
  \int_{r_1}^{r_2} \delta_k\;  \diff r = - \coupling - \glitch.
\end{equation}

Moreover, following \citet{Unno1989} $\delta_k$ can be express as a function of the g phase: $\int_{r_1}^{r_2} \delta_k\;  \diff r = \Phaseg$. By using Eq. (\ref{PhaseG_obs}) we can therefore write:

\begin{equation}\label{Relation_deltapi_buoyancyradius}
  \pi\left(\frac{1}{\nu\deltapi} - \eps_g \right) = - \coupling - \glitch.
\end{equation}

After substituting the coupling phase $\coupling$ in this equation and using Eq. (\ref{PhaseP_obs}) to express the p-wave phase, we find the following relation:

\begin{equation}\label{modes_GlitchG}
  \nu = \nu_{n,\ell=1} + \frac{\Dnu}{\pi} \arctan \left( q \tan \left[\frac{\pi}{\nu\deltapi} - \eps_g + \glitch_G \right] \right),
\end{equation}
which is close from the expression derived by \citet{2012A&A...540A.143M}, having only the addition of the glitch phase.

For the case of a glitch modeled by a Gaussian-like function, Cunha et al. (in preparation) have demonstrated that the glitch phase $\glitch = \glitch_G$ is defined by:

\begin{equation}\label{GlitchG_description}
  \glitch_G = \arccot \left[\frac{2\pi\nu}{a_G \e^{-\frac{\width^{2}}{8(\pi\nu)^{2}}}\sin^{2}(\beta_2)}  - \frac{\cos(\beta_1)}{2\sin^{2}(\beta_2)} \right],
\end{equation}
where $\beta_1 = 2\frac{\Period}{2\pi\nu} + 2\coupling + 2\varepsilon$, $\beta_2 = \frac{\Period}{2\pi\nu} + \frac{\pi}{4} + \coupling + \varepsilon$ and $a_G = \Amp_G \left(\frac{r^{*}}{\sqrt{2}\Brunt_0^{*}} \right)^{1/2}$. $\varepsilon$ is a phase value and $\Brunt_0^{*}$ corresponds to the glitch-free buoyancy frequency at the glitch position in radius. It has to be noted that this formulation is only valid for a glitch situated in the outer cavity, with $\Period/\rescavity > 0.5$. The boundary conditions on the left and right side of the cavity are indeed not the same when p- and g-waves coupling happens.
\newline

With Eq. (\ref{modes_GlitchG}) and Eq. (\ref{GlitchG_description}) it is possible to derive the mixed-mode frequencies in the presence of a glitch. It has to be noted that this expression is only valid for a discontinuity described as a Gaussian-like function. Other discontinuity shapes can be present in stellar interiors, like in some stellar models with discontinuities appearing as discontinuous step-functions \citep{2008MNRAS.386.1487M,2015MNRAS.453.2290B}. In this case we model the glitch by the following  function:

\begin{equation}
  \Brunt = \left\{
       \begin{array}{l}
           \Brunt \ind{in} \ \ \ \mathrm{for}  \ \ \ r<r^{*}
           \\
           \Brunt \ind{out} \ \ \ \mathrm{for}  \ \ \ r>r^{*}
       \end{array}
   \right.
   \label{Equation:step_function}
\end{equation}
with $\Brunt$ varying by $\Delta\Brunt = \Brunt \ind{in} - \Brunt \ind{out}$. The glitch is therefore characterized by two parameters: its amplitude $\Amp_{ST} = [\Brunt \ind{in}/\Brunt \ind{out}]_{r^{*}} - 1$ and its radial position $r^{*}$. 

For this specific case, it has been shown that the glitch phase $\glitch = \glitch_{ST}$ is defined by (Cunha et al., in preparation):

\begin{equation}
   \glitch_{ST} = \arccot \left[ - \frac{2 + 2\Amp_{ST}\cos^{2}(\beta_2)}{ \Amp_{ST}\cos(\beta_1)} \right],
   \label{GlitchST_description_inner}
\end{equation}
if $\Period/\rescavity < 0.5$, which correspond to a glitch located in the inner half of the cavity (measured in terms of buoyancy radius), while

\begin{equation}
   \glitch_{ST} = \arccot \left[ - \frac{2 + 2\Amp_{ST}\sin^{2}(\beta_2)}{ \Amp_{ST}\cos(\beta_1)} \right],
   \label{GlitchST_description_outer}
\end{equation}
if $\Period/\rescavity > 0.5$, which correspond to a glitch located in the outer half of the cavity.

\subsection{Glitch influence on period echelle diagram\label{Glitch_influence}}

In order to evaluate the impact of the glitch parameters on the red-giant mixed-mode pattern, we computed the mixed-mode frequencies for a typical red-giant star using Eq. (\ref{modes_GlitchG}) with different glitch parameters and description. We used as an input, the global seismic parameters that were measured for the star KIC002692629. This red-giant branch star has a large separation and frequency of maximum oscillation equal, respectively, to $\Dnu = 16.23 \mu$Hz and $\numax = 207.37\mu$Hz. We settled the gravity-mode period spacing and coupling parameter to, respectively, $\deltapi = 85$s and $q = 0.13$ which correspond to typical values of these parameters for this kind of star \citep{2016A&A...588A..87V,2017A&A...600A...1M}. The frequencies of pure $\ell = 1$ pressure modes were determined by the use of the universal pattern \citep{2011A&A...525L...9M} which describes the pure p-mode eigenfrequency pattern. 

To isolate the effect of the glitch on the mixed-mode pattern, we modified the mode frequencies following the stretching technique as described in Section 2.2 of \citet{2016A&A...588A..87V}. In practice, each frequency is modified according to the difference expressed by the ratio between the mixed-mode period spacing (which corresponds to the difference between mixed-modes of consecutive radial order) and the pure gravity mode period-spacing ($\deltapi$). The stretched periods are therefore represented by a new variable: $\taufreq$. It corresponds to the mode periods for which the influence of the coupling between gravity and pressure waves has been removed. Following this statement, if the original mixed-mode pattern had no deviations from the analytic mixed-mode pattern \citep{Unno1989,2012A&A...540A.143M} the stretched mixed-mode frequencies would be regularly spaced in period. Since we initially introduce a glitch in our computed mixed-mode pattern, this will not be the case. However, using the stretching technique allows us to isolate the glitch influence and evaluate the impact of its different characteristics. 
\newline

\begin{figure*}[ht]
    \centering
	\includegraphics[width=0.28\linewidth]{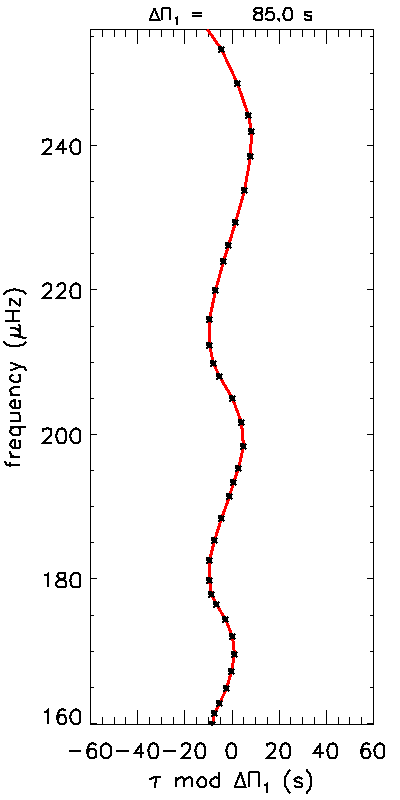}
	\includegraphics[width=0.28\linewidth]{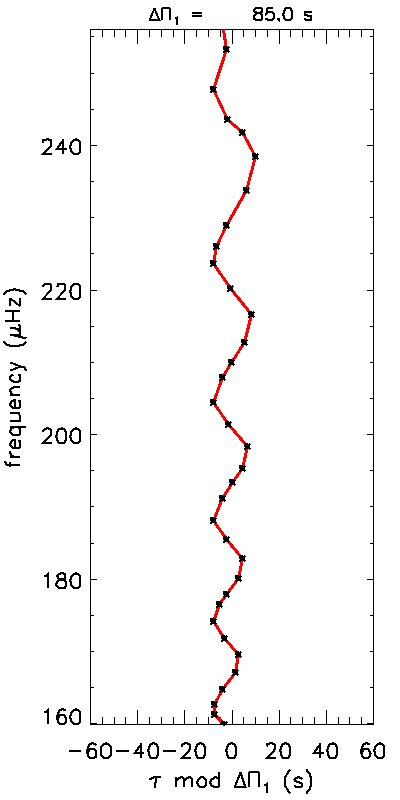}
	\includegraphics[width=0.28\linewidth]{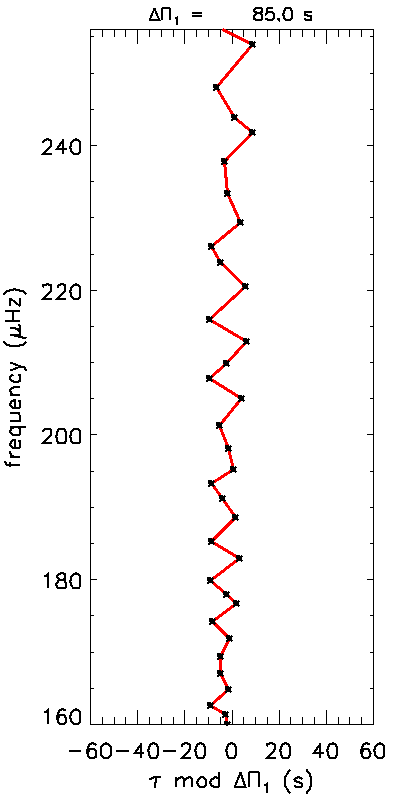}
	\caption{Stretched period echelle diagram of computed asymptotic mixed modes for which a buoyancy glitch modeled by a Gaussian with different radial positions in the resonant cavity were included. The glitch amplitude and phase was settled to, respectively, $\Amp_G = 0.005$ and $\varepsilon = 0.0$. From left to right, the glitch radial position corresponds to a ratio between the buoyancy depth and the total buoyancy radius ($\Period/\rescavity$) equal to: $0.10$, $0.20$ and $0.45$ respectively.}
	\label{fig:Fig_width}
\end{figure*}

The results for a glitch modeled by a Gaussian with different discontinuity positions in the gravity waves resonant cavity can be seen in Fig. \ref{fig:Fig_width}, represented as a stretched period echelle diagram. With this representation, if there is no deviations from the asymptotic mixed-mode pattern, we expect to see the mode aligned on a straight line. Here, the glitch influence appears clearly as a periodic modulation in the mixed-mode frequencies around the expected mixed-mode pattern as shown in previous works \citep{2015ApJ...805..127C,2015A&A...584A..50M}. As can be seen on Fig. \ref{fig:Fig_width}, the period of the modulation is affected by the glitch position inside the resonant cavity ($r^{*}$). A discontinuity present near the extremity of the resonant cavity induces a modulation with long period while a discontinuity close to the center of the resonant cavity results in a short-period modulation. The width of the discontinuity ($\width$) affects the amplitude of the modulation: it produces a decrease of the modulation amplitude with decreasing mode frequencies. The wider the glitch is, the stronger this decrease will be. Finally, the amplitude ($\Amp_G$) and phase ($\varepsilon$) of the discontinuity affect, respectively, the modulation amplitude and phase.
\newline

\begin{figure}
	\centering
	\includegraphics[width=0.45\linewidth]{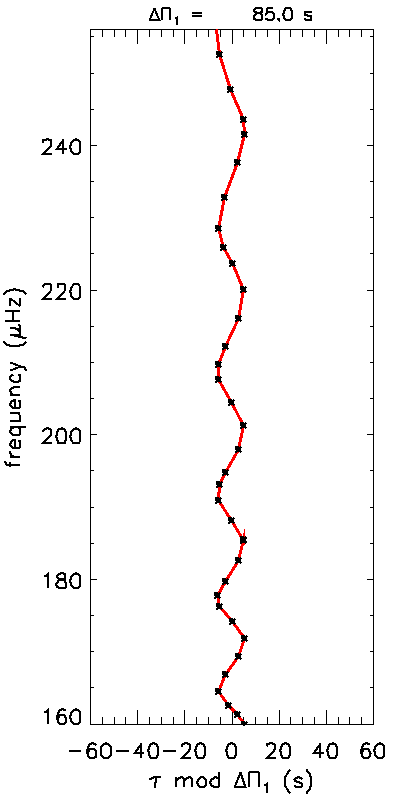}
	\caption{Stretched period echelle diagram of computed asymptotic mixed modes for which a buoyancy glitch modeled by a step-function was included. The glitch amplitude, phase and radial position was settled to, respectively, $\Amp_{ST} = 0.005$, $\varepsilon = 0.0$ and $\Period/\rescavity = 0.20$.}
	\label{fig:Fig_step}
\end{figure}

The impact of a discontinuity modeled by a step-function is very similar, as can be seen on Fig. \ref{fig:Fig_step}. However, since the shape of the discontinuity does not have a specific width, the amplitude of the observed modulation will have no variation as a function of the frequency.

\section{Characterization of the discontinuities}

In this section, the analytical expression providing the mixed-mode frequencies in the presence of a glitch in the stellar core will be used to characterize the deviations from the classical mixed-mode pattern in observed red-giant star spectra.

\subsection{Data selection}

Long-cadence data from $\Kepler$ up to the quarter Q$17$ were used, which correspond to $44$ months of photometric observations. We focused our sample on the $6111$ stars for which the signal-to-noise ratio was sufficient to determine their evolutionary status, following \citet{2016A&A...588A..87V}. Among them, we selected a few stars that show clear periodic deviations from the classical mixed-mode pattern as revealed in the period echelle diagram. We checked that these deviations do not correspond to the signature expected for core rotation. The stars showing these features were all belonging to the clump.

\subsection{Measurement of the mixed-mode frequencies}

To identify the mixed-mode frequency positions in the spectra, the measurement of the seismic global parameters is essential. We used the envelope autocorrelation function \citep{2009A&A...508..877M} in order to obtain estimates of the values of $\Dnu$, $\numax$ and the parameters of the Gaussian envelope power excess produced by the oscillations. After this, the $\Dnu$ values were refined by using the universal pattern \citep{2011A&A...525L...9M} in order to enhance the accuracy of the determination of $\Dnu$ and to precisely locate the different oscillation modes. 

In a second step, for a more precise estimation of the global seismic parameters, we performed a global fit of the background and the power following the bayesian fitting technique described by \citet{2015A&A...579A..83C}. We used the $\numax$ and the Gaussian envelope power excess parameters deduced from the previous measurement as priors for this fit. In order to obtain realistic priors for the background parameters, we used the scaling relations between these parameters and $\numax$ as obtained by \citet{2014A&A...570A..41K}. 

Following this, we also measured the $\deltapi$ and $q$ parameters using the automated method described in \citet{2016A&A...588A..87V}. We checked afterwards that the results given by the automated measurement were consistent and that there was no other possible solution. 

Since we are only interested in the mixed-mode characteristics, we selected the part of the spectra near the $\nu_{n,\ell=1}$ mode frequencies where they mostly appear. In practice, it consists on avoiding the frequency regions where radial and quadrupole modes are present. Since the positions of these modes are known through the red-giant universal pattern, the frequency selection only depends on the value of the large separation: we keep part of the spectrum with a second-order reduced frequency $\red$ verifying:

\begin{equation}\label{eqt-conditions}
    \red
     = {\nu \over \Dnu}
     -\left(\np+\epsilonp +{\alpha\over 2} (\np-\nmax)^2 \right)  \in [0.06,0.80]
     ,
\end{equation}
where $\epsilonp$ is the pressure asymptotic offset, $\np$ the pressure radial order, $\nmax = \numax/\Dnu-\epsilonp$ is the non-integer order at the frequency $\numax$ of maximum oscillation signal, and $\alpha$ is a term corresponding to the second order of the asymptotic expansion \citep{2013A&A...550A.126M}.

In order to identify the different mixed modes present in each portions of the spectra, we smoothed the power density spectrum with a low-pass filter with a width equal to $\Dnu/100$ and located the local maxima. These local peaks are considered to be significant if their heights exceed a threshold corresponding to the rejection of the pure noise hypothesis with a confidence level of $99.9\%$. Since only $\ell = 3$ modes with very low visibility can be present at those frequencies, all the significant peaks are assumed to be $\ell = 1$ mixed modes. 

\begin{figure}
	\centering
	\includegraphics[width=1.00\linewidth]{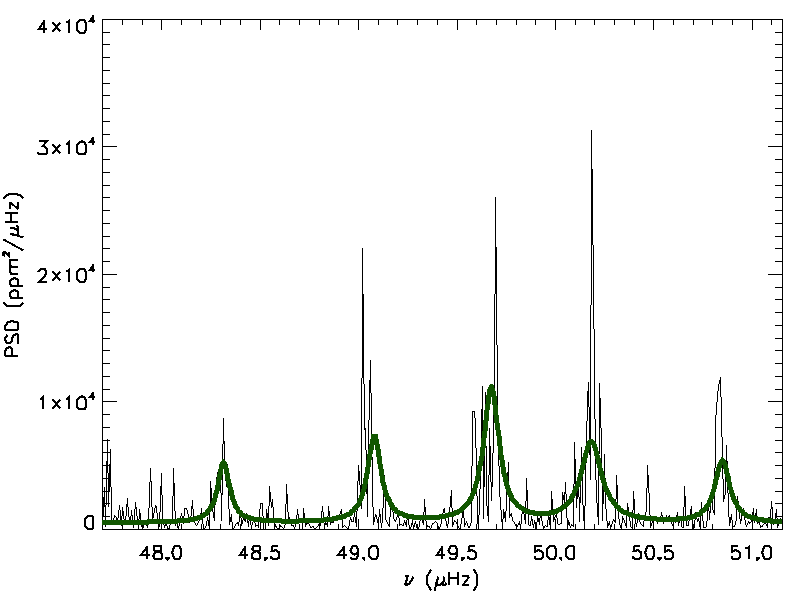}
	\caption{Power density spectrum for the star KIC1995859 (black line) as a function of frequency. The fit of the modes is shown with the solid green line. Here it consists of several Lorentzians, since all modes are identified as being resolved, fitted over the previously adjusted background.}
	\label{fig:mode_fitting}
\end{figure}

Finally, we fit all the modes that were identified on each portion of the spectra using the bayesian method described in Part $2.2$ of \citet{2015A&A...579A..83C}. Two types of peak profiles were used to fit the modes following the fact that resolved and unresolved mixed modes both exists in those spectra. Resolved peaks are modes for which the observing time is higher or equivalent to the mode lifetime and unresolved peaks are modes for which the mode lifetime is largely higher than the observing time. In the former case, the oscillation peak profile is represented as a sinus cardinal function \citep{2004SoPh..220..137C} otherwise, the adopted profile is that of a Lorentzian \citep[e.g. ][]{1990ApJ...364..699A}. In order to know if the different peaks correspond to resolved modes or not, we analyzed the number of frequency bins, which belongs to the peaks, that are above $8$ times the background level, thus corresponding to the presence of a signal with a confidence level of $99.9\%$. If there is more than one frequency bin that reach this level for a specific peak, then the corresponding mode is considered as being resolved otherwise it is identified as unresolved. An example of the performed adjustment is shown in Fig. \ref{fig:mode_fitting} for one portion of the star spectrum KIC$1995859$. The fit of the different portions of the spectra following these criteria allowed us to extract the frequencies of all significant modes for the selected objects.

\subsection{Gravity-mode glitches identification and characterization}

\begin{figure}
	\centering
	\includegraphics[width=0.48\linewidth]{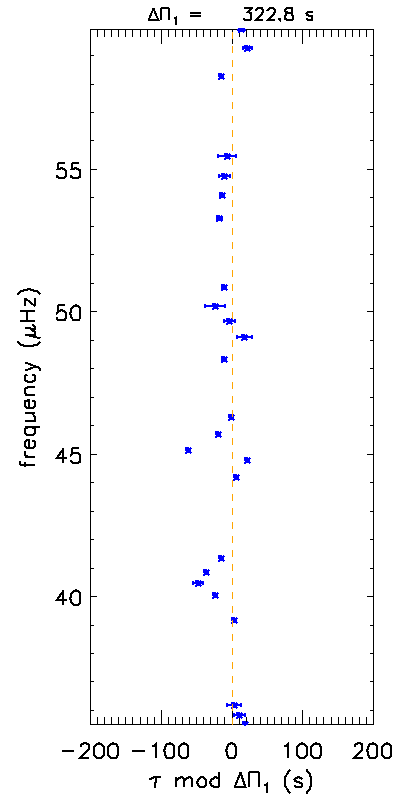}
	\includegraphics[width=0.48\linewidth]{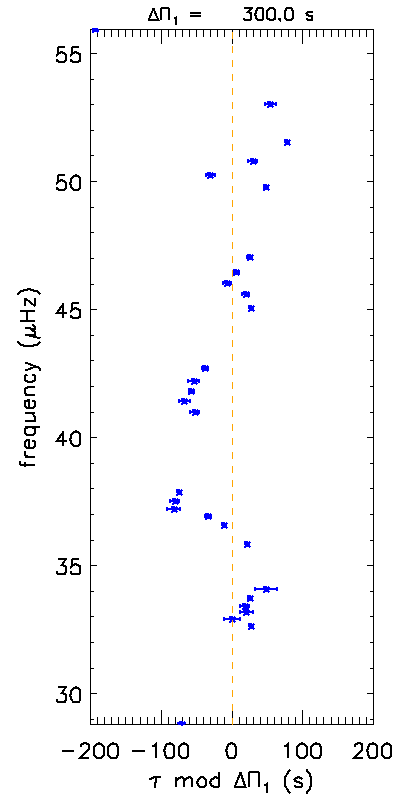}
	\caption{Stretched period echelle diagram of the stars KIC$1995859$ (left) and KIC$9332840$ (right). Error bars correspond to the $1-\sigma$ uncertainties. The dashed orange line indicates the stretched period reference if there was no shift due to glitches from the asymptotic development.}
	\label{fig:Stretched_diagram}
\end{figure}

In order to precisely characterize the glitches, we performed the stretching of the frequencies as described in Section \ref{Glitch_influence}. As stated before, it allows us to isolate the glitch influence from the coupling influence on the mixed-mode pattern. The results for the stars KIC$1995859$ and KIC$9332840$ are presented in Fig. \ref{fig:Stretched_diagram}. For the star KIC$1995859$, the modes exhibit a single nearly vertical ridge, showing that there is no obvious regular deviations. However, it is not the case for other stars like KIC$9332840$ for which a long modulation is observed as what was observed for the asymptotic development in Section \ref{Glitch_influence}.

\begin{figure*}
	\centering
	\includegraphics[width=0.30\linewidth]{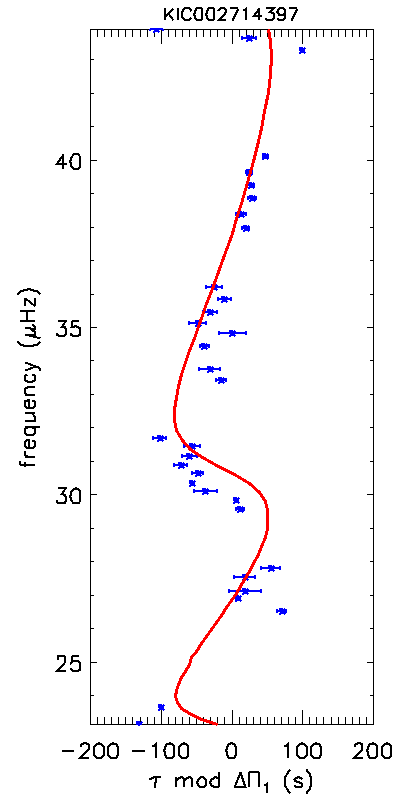}
	\includegraphics[width=0.30\linewidth]{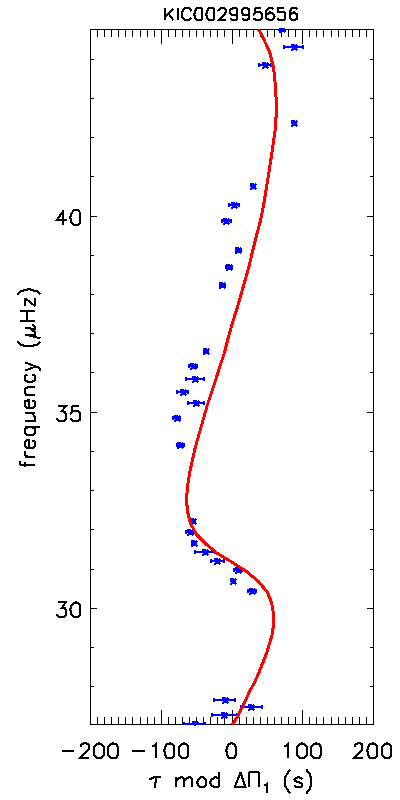}
	\includegraphics[width=0.30\linewidth]{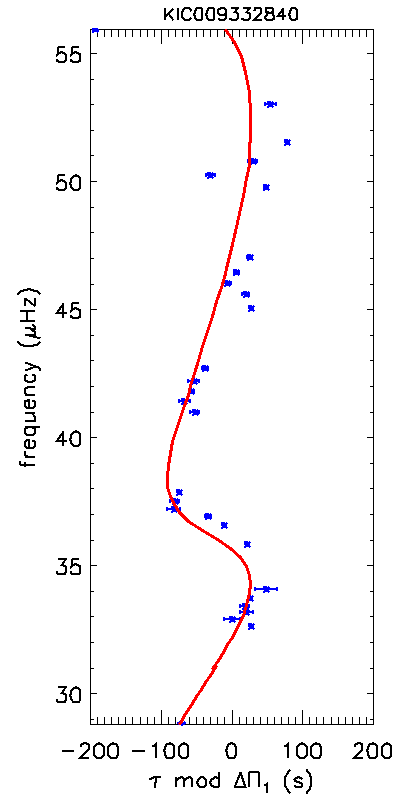}
	\caption{Stretched period echelle diagram for the stars KIC$2714397$ (left), KIC$2995656$, and KIC$9332840$ (right). Error bars correspond to the $1-\sigma$ uncertainties. The fit of the observed modulation is shown with the solid red line.}
	\label{fig:Stretched_diagram_fit}
\end{figure*}

Among the clump stars we considered in the sample, we could identified $15$ for which clear regular modulation was present. In order to characterize the parameters of the modulation, we performed a fit based on Eq. (\ref{modes_GlitchG}). The glitch model we used is the one using a step-function (Eq. (\ref{GlitchST_description_inner}) and Eq. (\ref{GlitchST_description_outer})) since this model has less free parameters. Moreover, the frequency range where we detect modes is too small to account of the amplitude variation with the frequency we expect to see for the glitch modeled by a Gaussian. The fit was realized with a $\chi^{2}$ estimator. The error bars were deduced through the inversion of the Hessian matrix. The fit was successful for $10$ stars, the results for three of them are shown on the Fig. \ref{fig:Stretched_diagram_fit}. The detailed results can be seen in Table \ref{tab:table_wide}.

\begin{table*}[t]
	\centering
	\caption{$\deltapi$, coupling parameter and glitch fitting results for the successfully fitted mixed-mode pattern of the stars we selected}
	\label{tab:table_wide}
	\begin{tabular*}{\linewidth}{c @{\extracolsep{\fill}} c c c c c}
	\noalign{\smallskip}\hline\hline\noalign{\smallskip}
	KIC number & $\deltapi$ (s) & q & $\Period/\rescavity$ & $\Amp_{ST}$ & $\varepsilon$\\
	\noalign{\smallskip}\hline\noalign{\smallskip}
	$1724879$ &  $293.8$ & $0.22$ &  $0.044\pm0.005$  & $2.09\pm0.12$ & $-2.34\pm0.03$ \\
	$1726211$ &  $322.0$ & $0.36$ &  $0.024\pm0.010$  & $2.09\pm0.47$ & $1.86\pm0.05$ \\
	$2156988$ &  $280.3$ & $0.31$ &  $0.009\pm0.025$  & $3.00\pm0.26$ & $-1.04\pm0.02$ \\
	$2303367$ &  $307.7$ & $0.31$ &  $0.034\pm0.025$  & $2.40\pm0.03$ & $-3.04\pm0.03$ \\
	$2583651$ &  $280.2$ & $0.29$ &  $0.016\pm0.005$  & $4.00\pm0.13$ & $-1.04\pm0.01$ \\
	$2714397$ &  $324.5$ & $0.39$ &  $0.034\pm0.025$  & $3.00\pm0.06$ & $2.36\pm0.01$ \\
	$2995656$ &  $306.1$ & $0.29$ &  $0.024\pm0.005$  & $2.90\pm0.18$ & $-0.34\pm0.02$ \\
	$3117024$ &  $248.0$ & $0.21$ &  $0.020\pm0.005$  & $3.00\pm0.32$ & $0.56\pm0.02$ \\
	$3946270$ &  $304.0$ & $0.39$ &  $0.036\pm0.010$  & $4.00\pm1.26$ & $-2.74\pm0.06$ \\
	$9332840$ &  $300.0$ & $0.24$ &  $0.030\pm0.005$  & $2.70\pm0.14$ & $-1.54\pm0.01$ \\
	\noalign{\smallskip}\hline
	\end{tabular*}
\end{table*}

\section{Discussion and conclusion}

As can be seen in Fig. \ref{fig:Stretched_diagram_fit}, the modulations we discovered for these clump stars have very long periods. Therefore, these periods correspond to a discontinuity located close to one of the extremities of the gravity waves resonant cavity. In red-giant clump stars, there are few discontinuities that could theoretically produce such a glitch. One of them is the region of hydrogen-burning shell. However, this region is situated, during the clump evolutionary states, at the middle of the gravity waves resonant cavity and is usually too broad to produce a firm discontinuity \citep{2015ApJ...805..127C}. The second discontinuity we can consider is the signature of the first dredge-up \citep{1967ARA&A...5..571I}. However, like for the hydrogen-burning shell, this discontinuity is situated closer to the middle of the resonant cavity in intermediate-mass clump stars and it disappears during the red-giant branch stellar evolution phase for low-mass stars \citep{2015ApJ...805..127C}. The last physical process that can produce a discontinuity is the influence of the convective core on the radiative part of the stellar core. There are indeed several physical processes induced by the development of a convective core during the clump phase like overshooting, penetrative convection \citep{2017MNRAS.469.4718B} or chemical mixing that will influence the base of the radiative region and produce discontinuities \citep{2015MNRAS.453.2290B}. More broadly, the influence of chemical discontinuities, present at the base of the radiative zone of the stars and produced by the chemical burning, on the gravity-mode pattern, has been theoretically predicted for other types of pulsators like $\gamma$-Doradus stars \citep[e.g. ][]{2008MNRAS.386.1487M} and hot B subdwarfs stars \citep{2013EPJWC..4304005C,2013ASPC..479..263C}. For the latter, recent observations have confirmed the theoretical predictions \citep{2018MNRAS.474.4709K}. These observations exhibit a modulation of the period echelle diagram that is close to what we observed in our study. Since these discontinuities are close to the inner limit of the gravity waves resonant cavity they could correspond to the ones we observed in this study. 
\newline

To conclude, we can say that we have conducted an analysis on the impact of core structural discontinuities on the mixed-mode pattern of red-giant stars. First, we performed a theoretical analysis, based on the asymptotic expansion developed by \citet[][2019]{2015ApJ...805..127C}, allowing us to predict the mixed-mode frequency positions in the presence of a discontinuity for different glitch models. We then selected a sample of red-giant stars and extracted the precise mixed-mode frequency positions in the oscillation spectra of those objects and fitted the theoretical glitch model on this observed mixed-mode pattern. We found a dozen red-giant clump stars exhibiting clear periodic frequency shifts in their mixed-mode pattern that correspond to the expected behavior of glitches. The characterization of the glitch parameters was possible for $10$ stars through the use of a fitting technique. We found that these observed glitches belong to a discontinuity situated near the extremity of the gravity waves resonant cavity and that they likely correspond to the influence of the stellar convective core. These results will have to be confirmed by increasing the sample size and improving the fitting of the glitch parameters in further studies.

\section*{Acknowledgments}
This work was supported by FCT - Fundação para a Ciência e a Tecnologia  through national
funds and by FEDER through COMPETE2020 - Programa Operacional Competitividade e
Internacionalização in the context of the grants: PTDC/FIS-AST/30389/2017 \& POCI-01-0145-FEDER-030389 and UID/FIS/04434/2013 \& POCI-01-0145-FEDER-007672.

MV also thanks St\'ephane Charpinet for fruitful discussions.

\bibliographystyle{phostproc}
\bibliography{Glitches_mvrard.bib}

\end{document}